\documentstyle[twocolumn,eqsecnum,aps,epsf,epsfig,subfigure]{revtex}
\newcommand{\PD}{\partial}
\newcommand{\be}{\begin{equation}}
\newcommand{\ee}{\end{equation}}
\newcommand{\bea}{\begin{eqnarray}}
\newcommand{\eea}{\end{eqnarray}}

\newcommand{\lton}{\mathrel{\lower.9ex
                  \hbox{$\stackrel{\displaystyle <}{\sim}$}}}

\begin{document}

\sloppy

\title{Hydrodynamics Near a Chiral Critical Point}
\author{K.\ Paech$^a$, H.\ St\"ocker$^a$, A.~Dumitru$^{a,b}$}
\address{$^a$Institut f\"ur Theoretische Physik, J.W.~Goethe Universit\"at,\\
Postfach 111932, D-60054 Frankfurt am Main, Germany}
\address{$^b$Department of Physics, Brookhaven National Laboratory, 
Upton, NY 11973-5000, USA}
\maketitle   


\begin{abstract}
We introduce a model for the real-time evolution of a relativistic
fluid of quarks coupled to non-equilibrium dynamics of the long wavelength
(classical) modes of the chiral condensate. We solve the equations of motion
numerically in 3+1 space-time dimensions. Starting
the evolution at high temperature in the symmetric phase, we study
dynamical trajectories that either cross the line of first-order
phase transitions or evolve through its critical endpoint.
For those cases, we predict the behavior of the azimuthal momentum asymmetry
for high-energy heavy-ion collisions at nonzero impact parameter.
\end{abstract}

\narrowtext      


\section{Introduction}
The hydrodynamical model is frequently employed to describe
multiparticle production processes in hadronic
collisions~\cite{Landau:gs}. In particular, it predicts characteristic
flow-signatures as ``fingerprints'' for non-trivial equations of state of
hot and dense matter~\cite{EoS}. Such equations of state
can occur when the effective
potential, obtained by integrating out some degrees of freedom,
exhibits features
characteristic of a phase transition in thermodynamics~\cite{LeeWick}.

More specifically,
if there exist two (or more) collective states with the same free energy but
separated by a barrier, then behavior characteristic of a first-order phase
transition may emerge. On the other hand, if no free-energy barrier exists, one
might expect resemblence to a second-order phase transition. 
This analogy of interacting quantum field theories with thermodynamics is
believed to have played an important role in the evolution of the early
universe~\cite{KolbTurner} and is currently being investigated in
accelerator experiments by colliding beams of protons and heavy
ions~\cite{HarrisMueller}. Classical energy flow and
hydrodynamic scaling behavior emerges in high-energy inclusive
processes~\cite{Landau:gs,Low:1977gq} and from the real-time evolution of some
quantum field theories~\cite{RTQFT}.

In this paper, we extend the hydrodynamical transport model such that phase
transitions related to the restoration (or breaking) of some global symmetry
can be studied dynamically. In particular, we focus on
chiral symmetry breaking at finite temperature~\cite{Pisarski:1984ms}, for
which we shall adopt a relatively simple and tractable phenomenological model,
i.e.\ the Gell-Mann and Levy model~\cite{Gell-Mann:np}.

It has been argued~\cite{Stephanov:1999zu,Halasz,OveDirk,Mocsy:2001za}
that the chiral phase transition for 
two massless quark flavors is second-order
at baryon-chemical potential $\mu_B=0$,
which then becomes a smooth crossover for small quark masses.
On the other hand, a first-order phase transition is predicted for small
temperature $T$ and large $\mu_B$. If, indeed, there is a smooth crossover for 
$\mu_B=0$ and high $T$, and a first-order transition for small $T$ and
high $\mu_B$, 
then the first-order phase transition line in the $(\mu_B,T)$ plane must end 
in a second-order critical point. This point was
estimated~\cite{Stephanov:1999zu} to be at $T\sim100$~MeV and $\mu_B\sim 
600$~MeV (see also~\cite{OveDirk,Mocsy:2001za}).
More recently, it has been attempted to determine the endpoint of
the line of first-order phase transitions from the lattice,
using $2+1$ quark flavors on $N_t=4$ lattices~\cite{Fodor:2001pe}
(see also~\cite{Allton:2002zi}). Those authors locate the critical
point at $T=160 \pm 3.5$~MeV and $\mu_B= 725 \pm 35$~MeV. Note, however,
that a reliable extrapolation to the continuum limit and to physical
pion mass has not been attempted so far.

There is an ongoing experimental effort to detect that chiral critical point
in heavy-ion collisions at high energies. Note that both, high temperature
{\em and} high baryon density are required to have dynamical trajectories
in heavy-ion collisions pass reasonably close by the critical point.
Some dynamical computations~\cite{joerg}
of the energy deposition and baryon stopping
process during the initial stage of head-on collisions of large nuclei
within semi-realistic multi-fluid dynamical models suggest
that the required conditions may be reached in the central region of
collisions at $E_{lab} \simeq 20-80$~AGeV on a fixed target, or in the
fragmentation regions of collisions at higher energies. 
However, to our knowledge there has been so far no attempt to describe
hydrodynamical expansion of the hot and dense droplet produced initially
for dynamical trajectories close to the critical point. This paper
represents an attempt in that direction.

\section{The Model}
In this section we shall present our model for the dynamics of a droplet
of quarks and antiquarks, starting at high temperature in a state with
(approximately) restored chiral symmetry, and evolving towards a state where
the symmetry is spontaneously broken. The quarks will be described as
a heat bath in local thermal equilibrium that evolves in 3+1 dimensions
according to the conservation laws for energy and momentum, i.e.\ relativistic
hydrodynamics. However, the ``fluid'' of quarks interacts locally with the
chiral fields, that is, they can exchange energy and momentum. In turn, the
(long wavelength modes of the) fields obey the classical equations of motion
which follow from the underlying Lagrangian in the presence of the quarks and
anti-quarks. Similar models for the dynamics of quarks coupled to chiral
fields were considered in the past. In~\cite{CsMi}, a background of freely
streaming quarks was assumed, and the classical evolution of the chiral fields
was discussed. More realistic dynamical descriptions for the quark medium
followed shortly, treating them as either a relativistic fluid~\cite{MPS},
as also envisaged here, or within classical Vlasov transport
theory~\cite{Abada,abada_aich}. Those studies focused on a second-order chiral
 phase
transition, or, in the presence of explicit symmetry breaking, on a smooth
crossover. However, it turns out that one can also address first-order
chiral phase transitions within the very same model, at least in a
phenomenological fashion, by chosing larger quark-field coupling
$g$~\cite{Scavenius:1999zc,ove_bub} (see below).
Integrating out the quarks then leads to an effective
potential exhibiting two degenerate states around $T_c$.

Here, we extend the previous work mentioned above, and at the same time shift
our focus somewhat. Namely, the early studies were mainly concerned with the
dynamical evolution of the long-wavelength chiral fields, and of classical
pion production; that is, they mainly adressed issues related to the
possiblity of forming ``Domains of Disoriented Chiral Condensates'' (DCC),
as suggested by Rajagopal and Wilczek, and others~\cite{DCC}, see 
also~\cite{Mocsy:2001za,MPS,Abada,Scavenius:1999zc}.
Our present work puts more
emphasis on the dynamics of the heat-bath of quarks, rather than on that of
the soft modes of the chiral field. We shall point out qualitative
changes in the classical energy-momentum flow of the ``fluid'' of quarks
in the proximity of a chiral critical point, rather
than look for ``rare phenomena'' like DCC formation.

Moreover, refs.~\cite{CsMi,MPS,Abada,abada_aich,Scavenius:1999zc} all employed the mean
field approximation for the chiral fields. Field fluctuations at the phase
transition were not considered. As an example, for the first-order phase
transition discussed in~\cite{Scavenius:1999zc} dynamical bubble nucleation
(``boiling'') could not be described, as it requires large coherent thermal
field fluctuations from the symmetry restored phase, over the barrier and into
the symmetry broken phase (see e.g.~\cite{PL_bub} for results of such
dynamical simulations, and~\cite{ove_bub} for a computation of bubble
nucleation rates from the linear sigma model). Thus, the main
improvement here is that we do include a dynamical treatment of field
fluctuations in the
vicinity of a critical point, and their influence on the dynamical evolution
of the quark fluid. 

On the technical side, going beyond the mean field approximation requires us
to introduce appropriate subtractions in all thermodynamical functions,
as explained in appendix~\ref{app_subtr}. Moreover, 
the coupled system of non-linear partial differential equations has to be
solved numerically in 3+1 dimensions, without imposing any space-time symmetry
assumptions (while~\cite{CsMi,MPS,Abada,Scavenius:1999zc} all simplified the
solution greatly by assuming special symmetries which essentially reduced
the problem to 0+1 or 1+1 space-time dimensions).
That is because fluctuations break
any space-time symmetry that may be obeyed by the mean field, as for example
spherical symmetry or symmetry under Lorentz boosts in a particular direction.

As mentioned in the introduction, physically the chiral critical point is
expected to occur for some specific values of temperature $T$ and
baryon-chemical potential $\mu_B$. To simplify the problem and its numerical
solution, however, here we rather choose to consider only locally
baryon symmetric matter, i.e.\ equal numbers of quarks and anti-quarks.
Instead, we can ``shift''
the critical point by varying the quark-field coupling constant $g$,
that is, by increasing or decreasing the vacuum mass of the constituent
quarks.
As explained in more detail below, large values for $g$ result in a
first-order phase transition, while small $g$ leads 
to a crossover. For the observables discussed in section~\ref{sec_results},
the qualitative difference between the two realizations of a chiral critical
point mentioned above should not matter much\footnote{However, we note that
there are indeed differences on the quantitative level. For example, for a
first-order phase transition in baryon dense matter the isentropic speed of
sound does not vanish in general at $T_c$~\cite{joerg,kerstin}, as it does for
zero net baryon charge, $\mu_B=0$.}. On the other hand, our simplified
treatment disables us from studying fluctuations of net baryon
charge~\cite{Bflucs}.

In section~\ref{sec_effpot} we discuss the effective potential ``seen'' by the
long wavelength modes of the chiral fields in the presence of a heat bath of
quarks and anti-quarks. Then, in~\ref{sec_Neq} we present our model
for the non-equilibrium dynamical treatment of field and fluid evolutions.
Some numerical algorithms and details are mentioned briefly in
section~\ref{numerics}. In section~\ref{sec_results} we present our
results and end with a summary and an outlook in section~\ref{outlook}.

\subsection{Effective Potential}  \label{sec_effpot}

As an effective theory of the chiral symmetry breaking
dynamics, we assume the linear $\sigma$-model coupled
to two flavors of quarks~\cite{Gell-Mann:np}:
\bea
{\cal L} &=&
 \overline{q}\,\left[i\gamma ^{\mu}\partial _{\mu}-g(\sigma +\gamma _{5}
 \vec{\tau} \cdot \vec{\pi} )\right]\, q\nonumber\\
&+& \frac{1}{2}\left (\partial _{\mu}\sigma \partial ^{\mu}\sigma + 
\partial _{\mu} \vec{\pi} \partial ^{\mu}\vec{\pi} \right)
- U(\sigma ,\vec{\pi})\quad.
\label{sigma}
\eea
The potential, which exhibits both spontaneously and explicitly broken 
chiral symmetry, is
\begin{equation} \label{T=0_potential}
U(\sigma ,\vec{\pi} )=\frac{\lambda ^{2}}{4}(\sigma ^{2}+\vec{\pi} ^{2} -
{\it v}^{2})^{2}-h_q\sigma -U_0\quad.
\end{equation}
Here $q$ is the constituent-quark field $q=(u,d)$. The
scalar field $\sigma$ and the pseudoscalar field $\vec{\pi} 
=(\pi_{1},\pi_{2},\pi_{3})$ together form a chiral field $\phi_a = 
(\sigma,\vec{\pi})$. The parameters of the Lagrangian are chosen such that 
chiral $SU_{L}(2) \otimes SU_{R}(2)$ symmetry is spontaneously broken in the 
vacuum.  The vacuum expectation values of the condensates are
$\langle\sigma\rangle
={\it f}_{\pi}$ and $\langle\vec{\pi}\rangle =0$, where ${\it f}_{\pi}=93$~MeV 
is the pion decay constant. The explicit symmetry breaking term is due to the 
non-zero pion mass and is determined by the PCAC relation, which gives
$h_q=f_{\pi}m_{\pi}^{2}$, where $m_{\pi}=138$~MeV.  
This leads to $v^{2}=f^{2}_{\pi}-{m^{2}_{\pi}}/{\lambda ^{2}}$.  The value of 
$\lambda^2 = 20$ leads to a $\sigma$-mass, 
$m^2_\sigma=2 \lambda^{2}f^{2}_{\pi}+m^{2}_{\pi}$, approximately
equal to 600~MeV.  In 
mean field theory, the purely bosonic part of this Lagrangian exhibits a 
second-order phase transition~\cite{Pisarski:1984ms} if 
the explicit symmetry breaking term, $h_q$, is dropped.  For $h_q\ne 0$, the 
transition becomes a smooth crossover from the phase with restored symmetry
to that of broken symmetry. The normalization constant $U_0$ is chosen such
that the potential energy vanishes in the ground state, that is
\be
U_0 = \frac{m_\pi^4}{4\lambda^2} - f_\pi^2 m_\pi^2~.
\ee

For $g>0$, the finite-temperature one-loop effective potential also 
includes a contribution from the quarks. In our phenomenological approach,
we shall consider the quarks as a heat bath in (local) thermal
equilibrium. Thus, it is possible to integrate them out to obtain the effective
potential for the chiral fields in the presence of that bath of quarks.
Consider a system of quarks and antiquarks in thermodynamical
equilibrium at temperature $T$ and in a volume ${\cal V}$.
The grand canonical partition function reads
\begin{equation}
{\cal Z}=
\int{\cal D}\bar{q}\, {\cal D}q\, {\cal D}\sigma\, {\cal D}\vec{\pi}\;
\exp\left[ \int_0^{1/T} d(it) \int_{\cal V} d^3 {\vec {x}}\, {\cal L} \right]
\,\, .
\end{equation}
(Anti-) Periodic boundary conditions for the (fermion) boson fields are
implied. In mean-field approximation the chiral fields in the
Lagrangian are replaced by their expectation values, which we
denote by $\sigma$ and $\vec{\pi}$.
Then, up to an overall normalization factor,
\begin{eqnarray}
{\cal Z} & = & {\cal N}_U
\int{\cal D}\bar{q}\, {\cal D}q\; 
\exp\Biggl\{  i \int_0^{1/T} dt \int_{\cal V} d^3 {\vec {x}} \nonumber\\
& &
 \bar{q} \left[ i\gamma ^{\mu}\partial _{\mu}-
g(\sigma +i\gamma _{5}\vec{\tau} \cdot \vec{\pi})\right]q
 \Biggr\} \nonumber \\
& = & {\cal N}_U \,
{\rm det}_p \Bigl\{ \bigl[ p_{\mu}\gamma^{\mu} -
g(\sigma +i\gamma_{5}\vec{\tau} \cdot \vec{\pi}) \bigr]/T\Bigr\}\,\, ,
\end{eqnarray}
where
\be
{\cal N}_U = \exp\left(-\frac{{\cal V}U(\sigma ,\vec{\pi} )}{T}\right)~.
\ee
Taking the logarithm of ${\cal Z}$, the determinant of the Dirac
operator can be evaluated in the standard fashion~\cite{kapustaFFT},
and we finally obtain the grand canonical potential
\begin{equation} \label{ptotsig}
-\frac{T}{{\cal V}}\log {\cal Z} = U + \widetilde{V}_q(T)~,
\end{equation}
where
\begin{equation} \label{pqq}
\widetilde{V}_{q}(T) = -d_{q} \int \frac{d^3{\vec{p}}}{(2\pi)^3}
\left\{E+T\, \log \left[1+e^{{-E}/{T}} \right] \right\}\,\, .
\end{equation}
Here, $d_q=24$ denotes the color-spin-isospin-baryon charge degeneracy of the 
quarks. For our purposes, the zero-temperature contribution to $\widetilde{V}_{q}$,
i.e.\ the first term in the integral in~(\ref{pqq}),
can be absorbed into $U$ via a standard renormalization of the bare
parameters $\lambda^2$ and $v^2$. The logarithmic dependence on the
renormalization scale is ignored in the following.

Adding the $T=0$ and the finite-$T$ contributions
defines our effective potential $V_{\rm eff}$:
\be \label{T>0_potential}
V_{\rm eff}(\phi_a,T) \equiv U(\phi_a) - d_q T
\int \frac{d^3p}{(2\pi)^3} \log\left(1+e^{-E/T}\right)\quad.
\ee
$V_{\rm eff}$ depends on the order parameter field through
the effective mass of the quarks, $m_q=g\sqrt{\phi^2}\equiv
g\sqrt{\sum_a \phi_a\phi_a}$, which enters into
the expression for the energy, $E = \sqrt{p^2 + g^2 \phi^2}$.

In principle, one should also integrate out short wavelength fluctuations of
$\phi$, which would lead to an additional contribution to the effective
potential~(\ref{T>0_potential}), see for
example~\cite{Mocsy:2001za,Mocsy:2002hv}. For the present phenomenological
analysis, however, the expression~(\ref{T>0_potential}) is already
sufficient in that it exhibits a critical endpoint for some
particular value of the coupling constant $g$. Since we do not expect the
simple model~(\ref{sigma}) to be quantitatively reliable anyway,
it is not unreasonable to employ the effective potential~(\ref{T>0_potential})
for a study of qualitative effects near a chiral critical point.

\begin{figure}[hp]
\centerline{\hbox{\epsfig{figure=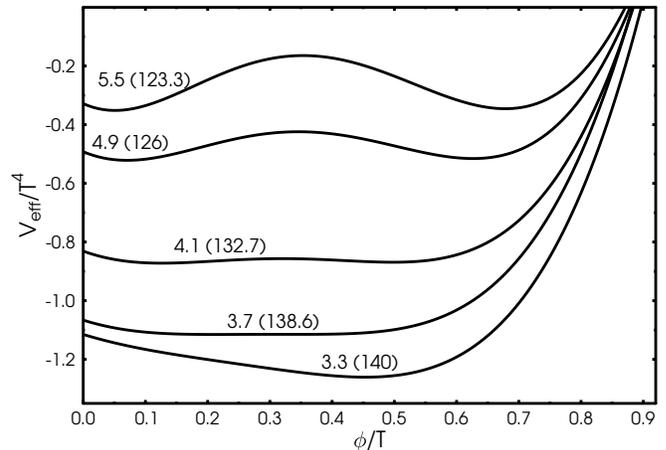,width=9cm}}}
\caption{The one-loop finite-$T$ effective potential as a function of
the scalar field $\sigma$ at $\vec\pi=0$.
The quark-field coupling constant $g$ is being varied.
The curves are labeled by the value for $g$ and by the temperature in MeV in
parantheses. The field self-coupling is chosen to be $\lambda^2=20$.}
\label{FigPotTc}
\end{figure} 
We now turn to a discussion of the shape of the effective
potential, cf.\ Fig.~\ref{FigPotTc}.
For sufficiently small $g$ one still finds the above-mentioned smooth 
transition between the two phases. At larger coupling to the quarks, however,
the effective potential exhibits a first-order phase
transition~\cite{Scavenius:1999zc}. Along the line of first-order transitions,
for temperatures near the critical temperature, $V_{\rm eff}$ displays
a local minimum $\sigma = \sigma_1(T) 
\simeq 0$ which is separated by a barrier 
from another local minimum at $\sigma=\sigma_2(T)>0$.
(There is another local minimum for negative $\sigma$ which is of higher energy
and does not concern us.)
These two minima are degenerate at $T=T_c$. For example,
$g=5.5$ leads to a critical temperature of
$T_c\simeq123.3$~MeV. Lowering the value of $g$ leads to a smaller barrier
between the two degenerate states. Also, $\sigma_1$ approaches $\sigma_2$,
i.e., the phase transition weakens, and moreover the spinodal temperature
approaches $T_c$~\cite{ove_bub}. At $g_c\simeq 3.7$,
finally, the barrier disappears, and so the latent heat vanishes.
This is the second-order critical point, where the potential about the
minimum is flat.

A different possibility of making the $\sigma$-meson much lighter
at $T_c$ than at $T=0$ is to reduce the self-coupling of the chiral
fields~\cite{GGP}, $\lambda^2$, rather than that to the quarks.
For example, one may choose $\lambda^2\simeq2.2$ with the pion decay
constant $f_\pi$ and vacuum mass $m_\pi$ fixed, such that $v^2=0$.
However, within the present model
this also reduces $T_c$ significantly, to less than
$100$~MeV~\cite{ove_bub}. Such low $T_c$ appear to be excluded
by present lattice QCD results~\cite{Laermann:1996xn}, and moreover would
generate too small thermal fluctuations in the heat bath. Therefore,
we keep $\lambda^2=20$ fixed throughout the manuscript.

\subsection{Coupled Dynamics of Fields and Fluid} \label{sec_Neq}
The classical equations of motion for the chiral
fields are
\bea
\partial_{\mu}\partial^{\mu}\sigma+ \frac{\delta U}{\delta\sigma}
  &=& -g\left<\bar{q}q\right>=-g\rho_s,  \nonumber\\
\partial_{\mu}\partial^{\mu}\vec{\pi}+\frac{\delta U}{\delta\vec\pi}
  &=& -g\left< \bar{q}\gamma_5 \vec\tau \,q\right>=-g\vec{\rho}_{ps},
\label{EulerLagrange}
\eea
where
\bea
\rho_s &=& \langle\overline{q}q\rangle
= g \sigma \, d_q\int \frac{{\rm d}^3 p}{(2\pi)^3}
 \frac{1}{E} f(p)\label{rho_s}~, \nonumber\\
\vec{\rho}_{ps} &=& \label{scal_dens}
\langle\overline{q}\gamma_5 {\vec{\tau}} q \rangle
= g \vec{\pi}\,  d_q\int \frac{{\rm d}^3 p}{(2\pi)^3}
 \frac{1}{E} f(p)
\eea
are the scalar and pseudoscalar densities
generated by the heat bath of quarks and anti-quarks, respectively.
The distribution of the quarks and anti-quarks in momentum
space is given by the Fermi-Dirac distribution.

The form~(\ref{EulerLagrange})
for the coupling to the heat bath of quarks can be derived from the Lagrange
density~(\ref{sigma}) in mean-field approximation.
The energy density of the quarks is given by
\be\label{qgphiq}
\left< {\cal H}_q \right> = \left<\bar{q}i\partial\!\!\!/ q\right> +
g\left<\overline{q} \sigma q\right>
+ g\left<\overline{q} \gamma_5\vec\tau\cdot\vec\pi q\right>~.
\ee
The finite-$T$ contribution to the equation of motion for the $\sigma$,
$\vec\pi$ fields
is obtained from the variation of the effective potential with respect to
$\sigma$ or $\vec\pi$, respectively:
\bea
g\left< \overline{q}q \right>  &=& 
\frac{\delta\left<{\cal H}_q\right>}
{\delta\sigma} = \frac{\delta\left<V_{\rm eff}-U\right>}
{\delta\sigma}~,\nonumber\\
g\left< \overline{q} \gamma_5\vec\tau\, q \right> &=&
\frac{\delta\left<{\cal H}_q\right>}
{\delta\vec\pi}= \frac{\delta\left<V_{\rm eff}-U\right>}
{\delta\vec\pi}~.
\eea
Applying this to the right hand side of equation~(\ref{T>0_potential})
yields the expressions for the scalar density and the pseudoscalar density
as given in~(\ref{rho_s}). 

As already mentioned above, we shall assume that the quarks constitute
a heat bath in local thermal equilibrium. Thus, their dynamical evolution is
determined by the local conservation laws for energy and momentum in
relativistic hydrodynamics. For simplicity, we shall further assume that
the stress-energy tensor of the quark fluid is of the ``perfect fluid''
form (corrections could in principle be taken into account in the future
along the lines discussed in~\cite{dissip}),
\be \label{idhyd_SET}
T^{\mu\nu} = (e+p)u^\mu u^\nu -pg^{\mu\nu}~.
\ee
Here,
\be
u^\mu \equiv \frac{T^{\mu\nu}u_\nu}{\sqrt{u_\sigma T^{\sigma\rho}
T_{\rho\alpha}u_\alpha}}
\ee
is the local four-velocity of the fluid. $g^{\mu\nu}={\rm diag}(1,-1,-1,-1)$
is our metric tensor, and so the line element is $d s^2 = 
d t^2 - d\vec{x}^{\,2}$.
The time $t$ is measured in the global restframe. Furthermore,
\bea
e(\phi,T) &=& \left<{\cal H}_q\right>~, \nonumber\\
p(\phi,T) &=& -V_{\rm eff}(\phi,T)+U(\phi_a)~, \label{qEoS}
\eea
denote the energy density and the pressure of the quarks at temperature
$T$. Note that we do not assume that the chiral fields are in equilibrium
with the heat bath of quarks. Hence, both $e$ and $p$ depend explicitly on
$\phi=\sqrt{\sigma^2+\vec{\pi}^2}$. Given an initial condition
on some space-like surface, $T^{\mu\nu}$ at any other causally
connected space-time point is determined by
\be \label{contEq_Q}
\PD_\mu T^{\mu\nu} = S^\nu~.
\ee
In the absence of interactions between the chiral fields and the quarks, the
source term $S^\nu$ vanishes, and the energy and the momentum of the quarks,
\be
(E,\vec{P}) = \int d\sigma_\mu T^{\mu \nu}
\ee
are conserved:
\be
u_\mu \PD^\mu E = u_\mu \PD^\mu P^i =0~.
\ee
With interactions turned on, this is obviously not true any longer. Rather,
the {\em total} energy and momentum of fluid plus fields is the conserved
quantity:
\be\label{source_field}
S^\nu = - \partial_\mu T^{\mu\nu}_\phi~.
\ee
The stress-energy tensor for the fields can be computed from the Lagrange
density in the standard fashion. The effective mass of the quarks,
$m_q = g\sqrt{\sum_a \phi_a^2}$, is already accounted for in the 
EoS for the quark fluid, see eqs.~(\ref{qEoS}). Thus, $T^{\mu\nu}_\phi$
is the stress-energy tensor of the chiral fields alone:
\bea
T^{\mu\nu}_\phi &=& \sum\limits_a
\frac{\partial \left<{\cal L}_\phi\right>}{\partial\left(\partial_\mu \phi_a
\right)} \partial^\nu \phi_a
-g^{\mu\nu} \left<{\cal L}_\phi\right> \nonumber\\
{\cal L}_\phi &=& \sum\limits_a
\frac{1}{2}(\partial_{\mu}\phi_a)( \partial^{\mu}\phi_a) - U~.
\eea
Its divergence is given by
\bea
-\partial_\mu T^{\mu\nu}_\phi &=& - \sum\limits_a
\left\{ \partial_\mu \partial^\mu \phi_a + 
\frac{\delta U}{\delta\phi_a}\right\} \PD^\nu\phi_a \nonumber\\
&=& g \rho_s \PD^\nu \sigma +g \vec{\rho}_{ps} \cdot \PD^\nu \vec\pi~.
\label{div_Tmunu_phi}
\eea
In the second step we made use of the
equations of motion~(\ref{EulerLagrange}).
This is the source term in the continuity equation~(\ref{contEq_Q}) for the
stress energy tensor of the quark fluid. A different derivation based on
moments of the classical Vlasov equation for the quarks was given
in~\cite{CsMi,MPS}, see also~\cite{Abada}.

We emphasize again that we employ eqs.~(\ref{EulerLagrange}) not only
to propagate the mean field through the transition but fluctuations as
well. The initial condition includes some generic ``primordial''
spectrum of fluctuations, see section~\ref{sec_IniCond}, which then
evolve in the effective potential generated by the matter fields,
i.e.\ the quarks. Near the critical point, those fluctuations
have small effective mass and ``spread out'' to probe the flat
effective potential. Since all field modes are coupled, effectively
the fluctuations act as a noise term in the
equation of motion for the mean field, similar to the familiar
Langevin dynamics~\cite{Langevin}; near the phase transition, however,
the noise is neither Gaussian (the effective potential is not
parabolic) nor Markovian (zero correlation length in time) nor white
(zero correlation length in space). Rather, correlation lengths and
$n$-point functions of the ``noise''
are governed by the dynamics of the fluctuations
in the effective potential generated by the fluid of quarks, including
also effects from the finite size and the relativistic expansion of the system.

\subsection{Numerics and Technical Details} \label{numerics}
We briefly describe how we solved the coupled
system~(\ref{EulerLagrange},\ref{contEq_Q}) of partial differential equations
in 3+1 space-time dimensions.

We follow the evolution on $t=const$ hypersurfaces, and in a fixed spatial
cube of volume $L^3$. We discretize three-space in that cube by introducing
a $160^3$
grid with a spacing of $\Delta x=0.2$~fm (thus, $L=160\times 0.2$~fm=32~fm).
On that grid, we solve the hyperbolic continuity equations of fluid dynamics
(\ref{contEq_Q}) using the so-called phoenical SHASTA flux-corrected transport
algorithm with simplified source treatment. It is described and tested in
detail in Ref.~\cite{Dirk_num}, and we refrain from a discussion here.
The time step was chosen as $\Delta t=0.4\Delta x$, as appropriate for the
SHASTA~\cite{Dirk_num,multi_fluid}.
We performed each time step in the standard fashion with $S^\nu=0$, then
subtracted the sources $S^\nu$ from the energy and momentum density in the
calculational frame (i.e.\ the global rest frame of the fluid). Finally,
from $T^{00}$, $T^{0i}$ and the EoS $p=p(e,\phi_a)$ we solved for the
velocity of
the local restframe (LRF) of each cell, and for the energy
density $e$ of the fluid in the LRF. Such a treatment of sources in the
continuity equations for energy and momentum proved to work well in
relativistic multi-fluid dynamics~\cite{joerg,multi_fluid},
where one encounters a
similar equation due to interactions between various fluids.
The boundary conditions at the edge of the computational grid are such that
the fluid simply streams out when reaching the boundary. This can be monitored
by checking the conservation of the total energy as a function of time.

Regarding the fields, we solve the classical equations of motion using
a staggered leap-frog algorithm with second-order accuracy in time, see
for example Ref.~\cite{numrec}. The non-linear
wave-equations~(\ref{EulerLagrange})
are split into two coupled first-order equations (in time) by considering
separately the field $\phi_a$ and its canonically conjugate field
$d\phi_a/dt$. For this algorithm, the time step for propagation of the fields
has to be chosen smaller than that for propagation of the fluid. In
practice, we found that $\Delta t=0.1\Delta x$ gave reasonable accuracy.
As the time step for propagation of the fluid was chosen to be four times
larger, the fields where propagated for four consecutive steps, with the fluid
kept ``frozen'', followed by one single step for the fluid.
Spatial gradients of the fields where set to zero at the edges of the
computational grid, rather than employing periodic boundary conditions. The
spatial derivatives in the equations of motion for the
fields~(\ref{EulerLagrange}) where discretized to second order accuracy in
the grid spacing to match the second order accuracy in time. We employed the
same small grid spacing $\Delta x=0.3$~fm as for the fluid. Although the
mean fields vary on a larger scale, it is important to allow for non-linear
amplification of harder fluctuations which can couple to the
soft modes and affect the dynamical relaxation to the vacuum state with
broken symmetry, see for example~\cite{par_res} and~\cite{Aarts:2000wi}.

For the fluid, one needs to introduce nine 3d spatial grids\footnote{At finite
baryon density, one would
need another grid for $\rho_B$. Also, inverting the Lorentz
transformation to obtain the LRF densities $e$ and $\rho_B$ from
$T^{00}$, $T^{0i}$ and the net baryon current
$J_B^\mu\equiv \rho_B u^\mu$
in each time step would require a three-dimensional root search for the
function $p=p(e,\rho_B,\phi)$.}, for
$e$, $p$, $T^{0\mu}$, and $\vec u\,$.
In addition, one needs two more 3d grids for
each field component, namely for $\phi_a$ and $d\phi_a/dt$. 
To save computational resources, we have therefore decided to neglect the
pion field altogether, i.e., to set it to zero everywhere in the forward
light cone. Thus, only the $\sigma$-field is considered in the actual
computations described below. However, this represents a minor restriction
only, since we focus here on the bulk evolution of the quark fluid
rather than on fluctuation observables or
coherent pion production from the decay
of a classical pion field.
(Coherent pion production is known to contribute
a small fraction of the total pion yield
only~\cite{MPS,Abada,Scavenius:1999zc,DCC,gavin}.
Incoherent particle production from the decay of the fluid by far dominates,
if no kinematic cuts are applied.)


\section{Results} \label{sec_results}

Having formulated our model, we now proceed to show some specific examples
of numerical solutions. In particular, we would like to examine whether the
dynamical evolution changes as one crosses the chiral critical point.

\subsection{Initial Condition} \label{sec_IniCond}
As an example, we employ the following initial conditions for the results
described below. Of course, one could employ more refined initial conditions
and ``tune'' them such as to reproduce various aspects of the final state,
which in principle could be compared to experimental data. At present, our more
modest goal is to illustrate qualitative effects originating from the
phase transition. The initial conditions are meant to provide
a semi-realistic parameterization of the hot fireball created in a high-energy
heavy-ion collision.

At time $t=0$, 
the distribution of energy density for the quarks is taken to be uniform
in the $z$-direction (with length $2l_z = 12$~fm) and ellipsoidal in the
$x-y$-plane:
\be
e(t=0,\vec{x}) = \left\{ 
\begin{array}{r@{\quad : \quad}l}
e_{\rm eq} & x^2b^2 +y^2a^2 < (ab)^2 \wedge z < l_z \\
0	   & x^2b^2 +y^2a^2 > (ab)^2 \vee z > l_z
\end{array}
\right. \quad ,
\ee
where
$e_{\rm eq}$ denotes the equilibrium value of the energy density
taken at a temperature of $T_i\approx 160$~MeV.
In our calculation we choose $a=r_A-\tilde{b}/2$ and 
$b=\sqrt{r_A^2-\tilde{b}^2/4}$, where $\tilde{b}$ denotes the
impact parameter of two nuclei with radius $r_{A}$ (below, we choose
$r_A=6.5$~fm and $\tilde{b}=6$~fm).
Thus, the ellipsoidal shape resembles the almond shaped overlap of
two colliding nuclei. The $\Theta$-function distribution of the
initial energy density is evidently somewhat unrealistic; a smoother
distribution with non-zero surface thickness would be more
realistic and perhaps affect the results somewhat.
However, as already mentioned above, here we do not aim at
quantitative fits to experimental data but at illustrating qualitative
effects related to the shape of the effective potential in the
transition region.

The collective longitudinal velocity of the fluid of quarks
is assumed to rise linearly with $z$: 
$v_z(t=0,\vec{x})\propto z/l_z \cdot v_{max}$, where $v_{max}=0.2$~.
The transverse components of $\vec{v}$ are set to zero at $t=0$.

Our initial conditions for the chiral fields are
\bea
\sigma(t=0,\vec{x}) &=& \delta\sigma(\vec{x}) + f_\pi + \nonumber\\
& & {}(- f_\pi + \sigma_{\rm eq}) 
\cdot \left[ 1 +
\exp\left(\frac{\tilde{r}-\tilde{R}}{\tilde{a}}\right)
\right]^{-1} \nonumber\\
& & {}\cdot \left[ 1 +
\exp\left(\frac{|z|-l_z}{\tilde{a}}\right)\right]^{-1}~,
\nonumber\\
\vec\pi(t=0,\vec{x}) &=& \delta\vec{\pi}(\vec{x})\, ,
\eea
where $\tilde{r}=\sqrt{x^2+y^2}$, 
\be
\tilde{R} = \left\{ 
\begin{array}{r@{\quad : \quad}l}
\displaystyle
\frac{ab\tilde{r}}{\sqrt{b^2x^2+a^2y^2}} & \tilde{r}\neq 0 \\
a          & \tilde{r} = 0
\end{array}
\right. \quad ,
\ee
and 
$\tilde{a}=0.3$~fm is the surface thickness of this Woods-Saxon like distribution.
Here $\sigma_{\rm eq}\approx0$ is the value of the $\sigma$ field 
corresponding to $e_{\rm eq}$.
Thus, the chiral condensate nearly vanishes at the center, where the energy
density of the quarks is large, and then quickly interpolates to $f_\pi$
where the matter density is low.

$\delta\sigma(\vec{x})$ and $\delta\vec{\pi}(\vec{x})$ represent the initial
random fluctuations of the fields. Our focus at this stage is on how
those ``primordial'' fluctuations evolve through the phase transition
(or crossover) and how they affect the hydrodynamic expansion of the
thermalized matter fields (the fluid). Thus, for a first qualitative
analysis we do not rely on additional physics input\footnote{For
example, one might assume thermalized primordial fluctuations, in
which case their distribution depends on the effective potential at
the temperature $T_i$. We have checked, however, that at high
temperature $V_{\rm eff}(\phi,T_i)$ looks rather similar for all values of $g$
considered here, regardless of whether later on the evolution proceeds through
the crossover, the critical endpoint or the first-order phase
transition regimes.} for the
primordial fluctuations but rather chose a generic Gaussian distribution,
\be
 \label{gaussdis}
P[\delta\phi_a] \propto 
\exp\left( - \delta\phi_a^2/2\left<\delta\phi_a^2\right>\right)~,
\ee
with the variance $\left<\delta\phi_a^2\right>$ left as a free parameter.
(Here, no summation over the index $a$ (internal space) is carried out.)
We performed simulations with $\surd\left<\delta\phi_0^2\right>\equiv
\surd\left<\delta\sigma^2\right>=v/3$.
Such fluctuations are large enough to probe the barrier of the effective
potential in case of a first order transition, or the flat region close
to the critical endpoint (Fig.~\ref{FigPotTc}). On the other hand, they are
sufficiently small to allow for a one-loop subtraction of their
contribution to the effective potential, as described in
appendix~\ref{app_subtr}.

The time derivatives of the fields where set to zero at $t=0$,
with no fluctuations. As already mentioned in section~\ref{numerics}, the
actual numerical computations described below were performed with 
$\delta\vec\pi\equiv0$, i.e.\ $\left<\delta\phi_{1,2,3}^2\right>=0$.

One must further take into account that the field fluctuations are correlated
over some space-like distance $\xi\simeq1$~fm. This is a physical scale which
is present in the initial conditions; if one simply picks random fluctuations
at each point of the grid, the correlation length will instead be given by the
artificial numerical discretization scale $\Delta x$.

In practice, a useful and simple procedure to implement physical correlations
in the initial condition is as follows. First, at each point of the grid one
samples the distribution~(\ref{gaussdis}) at random. Then, one smoothly sweeps
a "sliding window'' of linear size $\xi=n\Delta x$ over the grid and averages
the fields:
\bea
\phi_a'(\vec x) &=& \frac{1}{n^3}  \nonumber\\
& & \hspace*{-1.3cm} \times \sum\limits_{i,j,k=0..n-1} \phi_a(\vec x +
 i\Delta x \vec{e}_1 + j\Delta x \vec{e}_2 + k\Delta x \vec{e}_3)~.
\eea
Here, $\vec{e}_1$, $\vec{e}_2$, $\vec{e}_3$ define a global cartesian
orthonormal basis, and $\vec x=1\vec{e}_1+1\vec{e}_2+1\vec{e}_3$, $1\vec{e}_1
+1\vec{e}_2+2\vec{e}_3$, $\cdots$, is the sequence of grid points.
Clearly, the above averaging procedure ``cools'' the fluctuations, in that
$\langle \delta\phi_a'^2\rangle \ne \langle \delta\phi_a^2\rangle$. 
In particular, in
the continuum limit $\Delta x\to0$ with $\xi=n\Delta x$ held constant,
one of course finds that $\langle \delta\phi_a'^2\rangle \to0$ for the
distribution~(\ref{gaussdis}). Therefore, in a final sweep over the grid one
has to rescale the fields at each grid point by
\be
\phi_a''(\vec x) = \phi_a'(\vec x) \sqrt\frac{\left<\delta\phi_a^2\right>}
{\left<\delta\phi_a'^2\right>}~.
\ee
This procedure leads to an initial field configuration that exhibits both the
proper physical correlation length and the desired fluctuations.
Moreover, it prevents the initial energy density from field gradients to
grow like $1/\Delta x^{\,2}$.

\subsection{Numerical Solution}
In the following we consider both
a first order phase transition corresponding to $g=5.5$, 
as well as the critical point at $g=3.7$.

\begin{figure}[h!]
\centering
\epsfig{figure=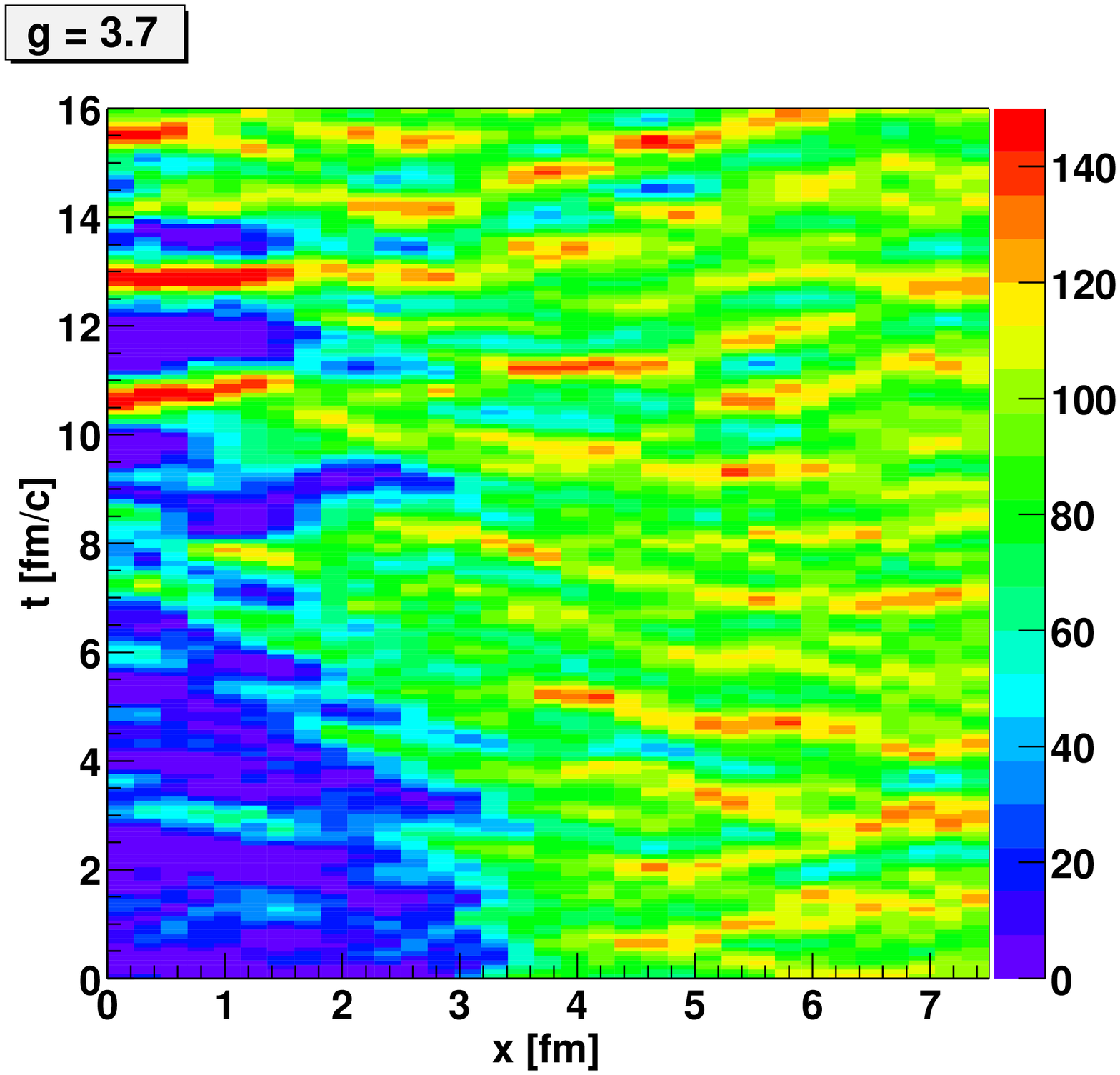,width=8cm}
\epsfig{figure=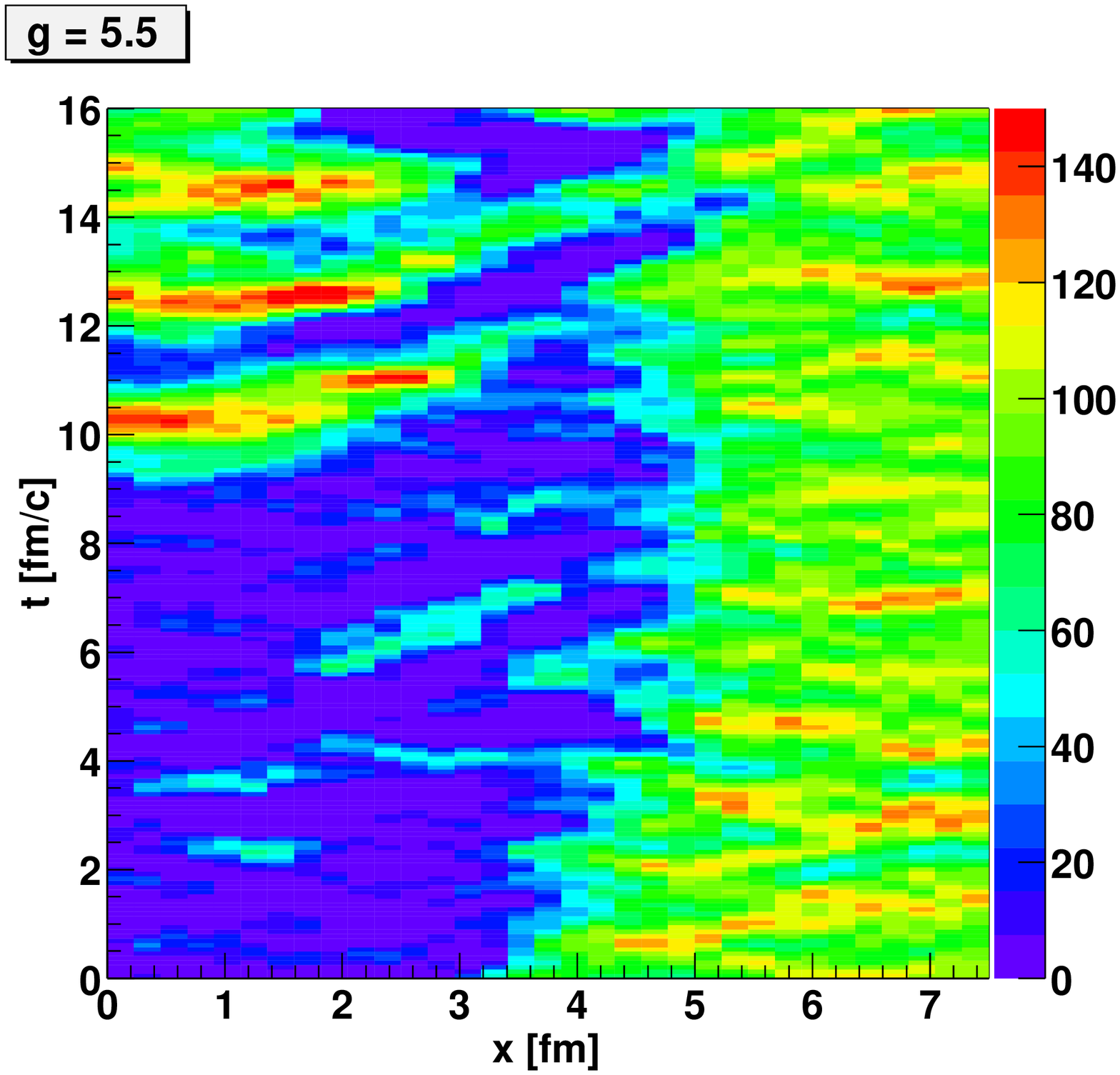,width=8cm}
\caption{Space-time evolution of the chiral field along the $x$-axis
at $y=z=0$. Dark regions correspond to small field amplitudes
(symmetric phase), light regions to large amplitudes (broken phase).
The scale on the right specifies the field amplitude in MeV.}
\label{compare_phix}
\end{figure}

\begin{figure}[h!]
\centering
\epsfig{figure=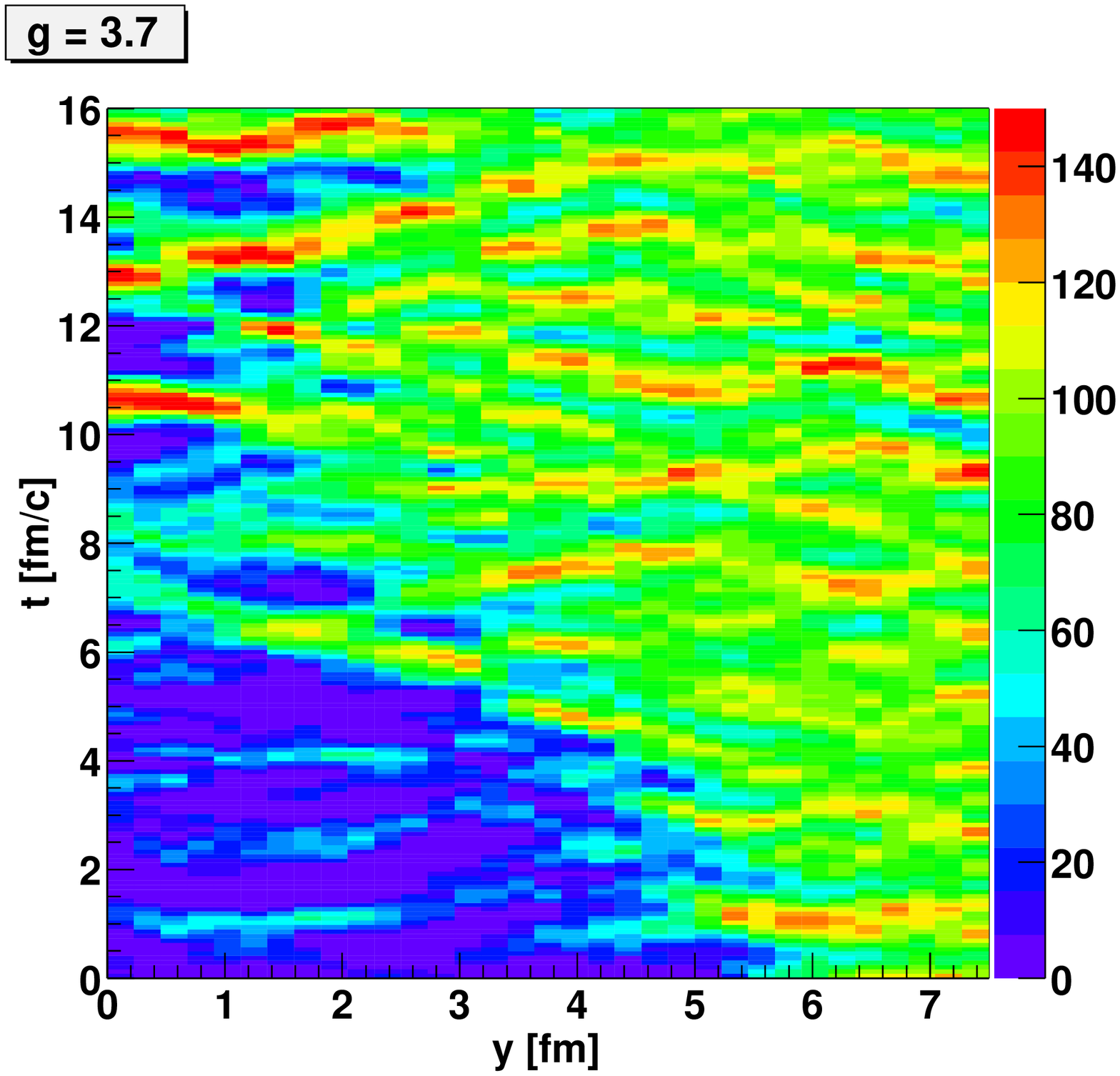,width=8cm}
\epsfig{figure=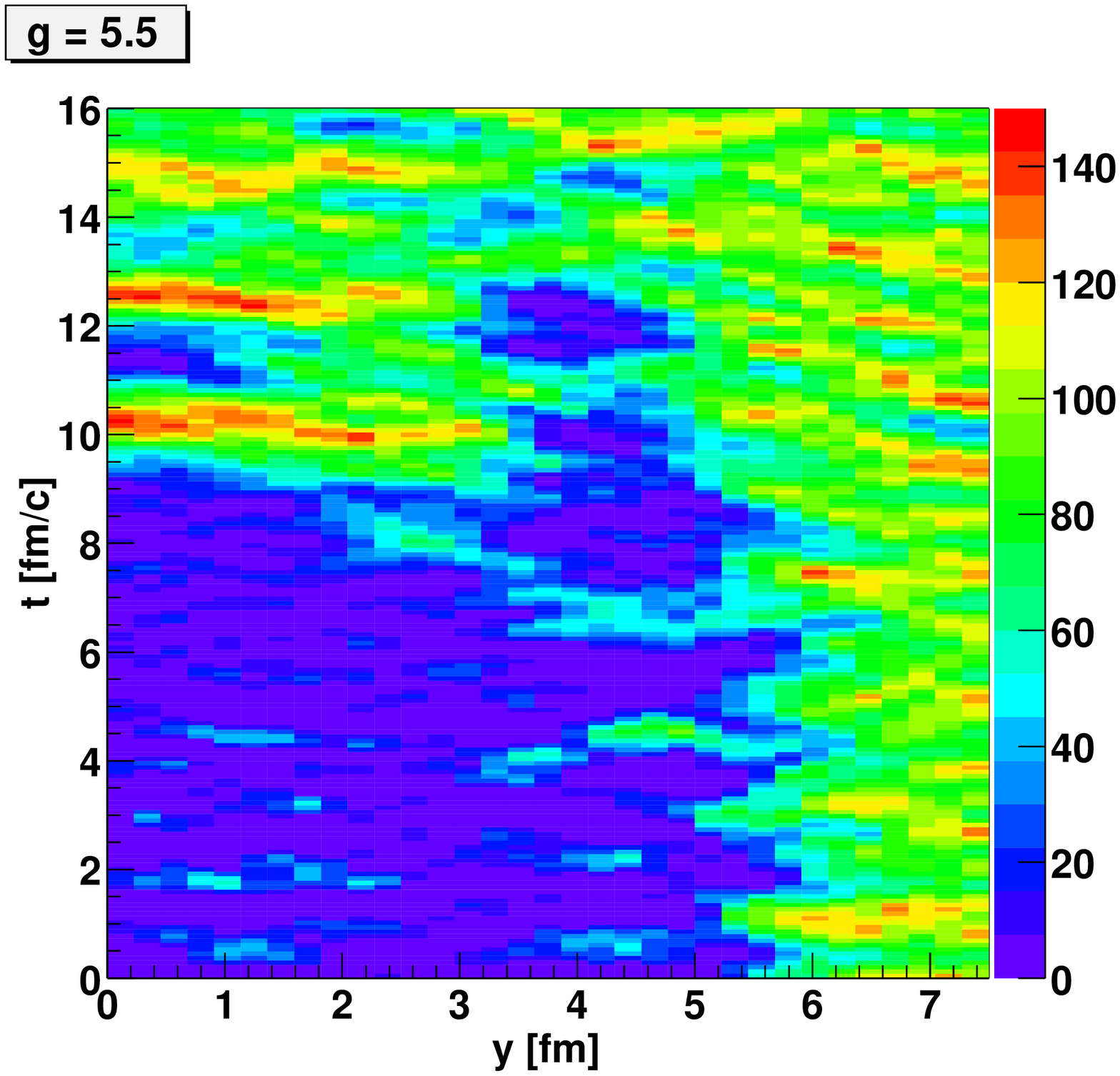,width=8cm}
\caption{Space-time evolution of the chiral field along the $y$-axis
at $x=z=0$.} 
\label{compare_phiy}
\end{figure}

Figs.~\ref{compare_phix}
and~\ref{compare_phiy} depict the time evolution of the
$\sigma$ field along the $x$-axis and the $y$-axis, respectively. 
At $t=0$ the field within the hot region has small amplitude,
corresponding to the chiral symmetry restored phase.
That region is surrounded by
the physical vacuum with $\langle\sigma\rangle=f_\pi=93$~MeV.

For the first order phase transition ($g=5.5$) a barrier separates the 
two degenerate minima of the effective potential (Fig.~\ref{FigPotTc}) 
at $T_c$.
Figs.~\ref{compare_phix} and~\ref{compare_phiy} show that this 
barrier leads to a rather well-defined surface
in coordinate space, separating the vacuum from the symmetric phase.
For the above-mentioned initial conditions, the fluctuations are not strong
enough for the field to easily overcome the barrier. Nevertheless, one can
observe dynamical fluctuations into the broken symmetry state, e.g.\
at $x\simeq1$~fm and $t\simeq2-4$~fm/c, which however collapse again.
Our dynamical results agree with previous arguments that
nucleation is a slow process on the time scale of heavy-ion collisions,
and so the Gibbs phase equilibrium is not established
dynamically~\cite{Scavenius:1999zc,ove_bub,PL_bub,Csorgo:dd}.
At time $t\simeq9$~fm the phase transition occurs spontaneously
``in an instant'', that is, on
a space-like surface which can clearly be seen in
Figs.~\ref{compare_phix} and~\ref{compare_phiy}.
A ``bubble'' of the symmetric phase survives at $x\simeq3$~fm for a long time.

\begin{figure}[h!]
\centering
\epsfig{figure=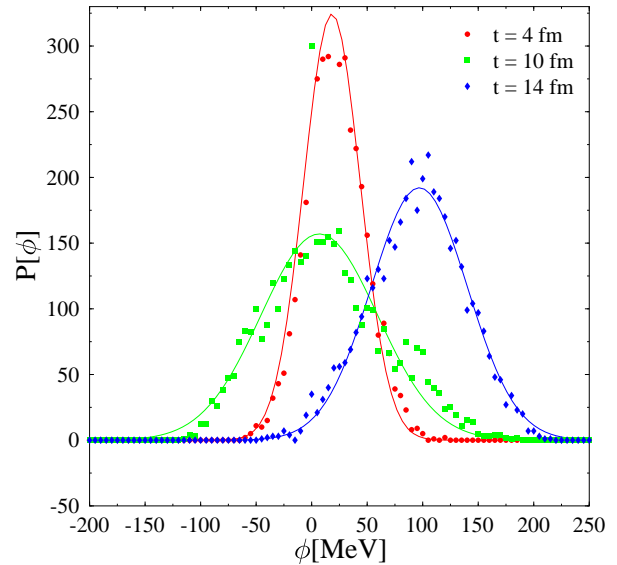,width=8cm}
\caption{Histogram of the field distribution at the center for 
three different times. Lines represent Gaussian fits with 
standard deviations 25~MeV (t=4~fm), 53~MeV (t=10~fm), and
43~MeV (t=14~fm).}
\label{histo}
\end{figure}

The picture is rather different for the transition at
the critical point, i.e.\ $g=3.7$. Here, the barrier between the
degenerate minima vanishes and the potential is flat. 
As is evident, there are no clear surfaces separating
either the vacuum from the center or high-density bubbles
(or ``droplets'') from their surrounding.
Due to the flatness of the potential, near the center
the field performs large-amplitude oscillations (from $\sigma\sim0$ to
$\sigma>f_\pi$) for a long time; they extend in space over distances
$\approx 1-3$~fm (e.g.\ at $t\approx 9$, 12, and 14~fm/c in
Fig.~\ref{compare_phix}), which is not much less than the initial size
of the hot region. Figure~\ref{histo} shows a histogram of the field
distribution at the center ($\sqrt{x^2+y^2+z^2}<2~fm$). One observes
that the distribution broadens from $t=4$~fm to  $t=10$~fm, and 
then narrows again at later times after the transition to the broken
phase occured. Also, notice that the distributions at $t=4$~fm and
$t=14$~fm are well described by Gaussians, i.e.\ the effective
potential is essentially parabolic, while for $t=10$~fm there are
visible deviations from a simple Gaussian. 

\begin{figure}[h!]
\centering
\epsfig{figure=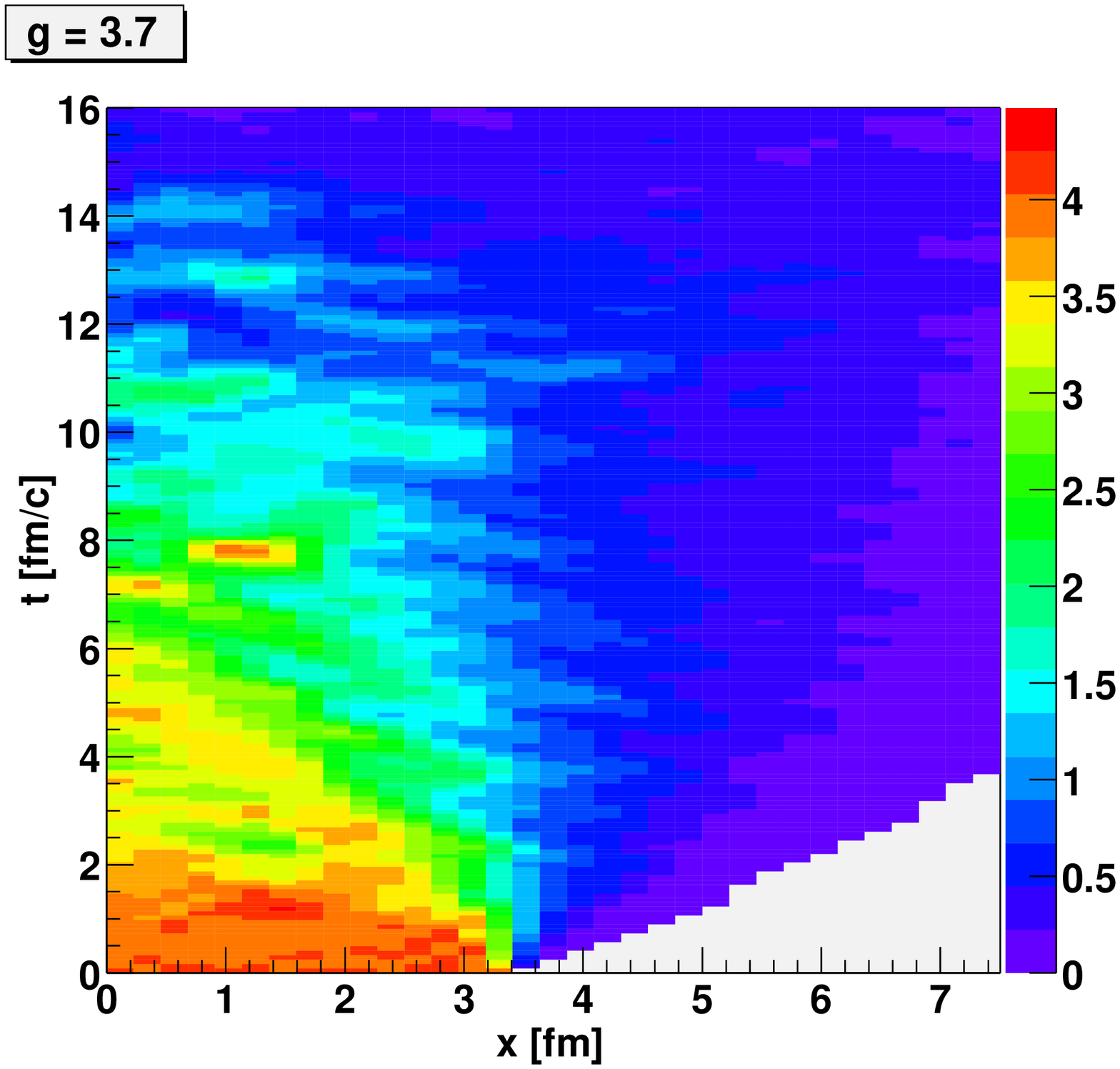,width=8cm}
\epsfig{figure=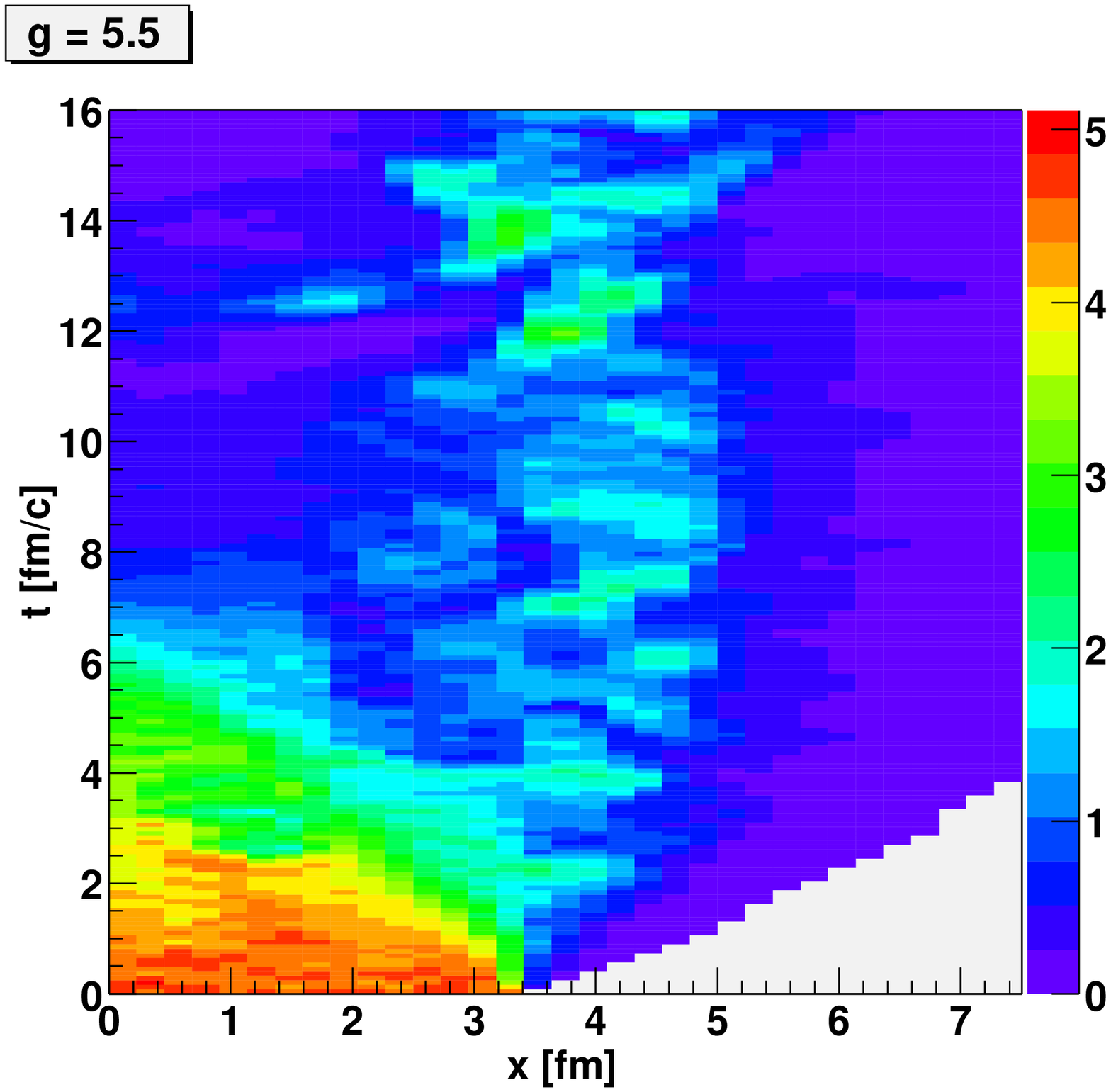,width=8cm}
\caption{Space-time evolution of the fluid energy density
along the $x$-axis at
$y=z=0$. The scale on the right specifies the energy density in units
of nuclear matter density $e_0\approx150$~MeV/fm$^3$.}
\label{compare_ex}
\end{figure}

\begin{figure}[h!]
\centering
\epsfig{figure=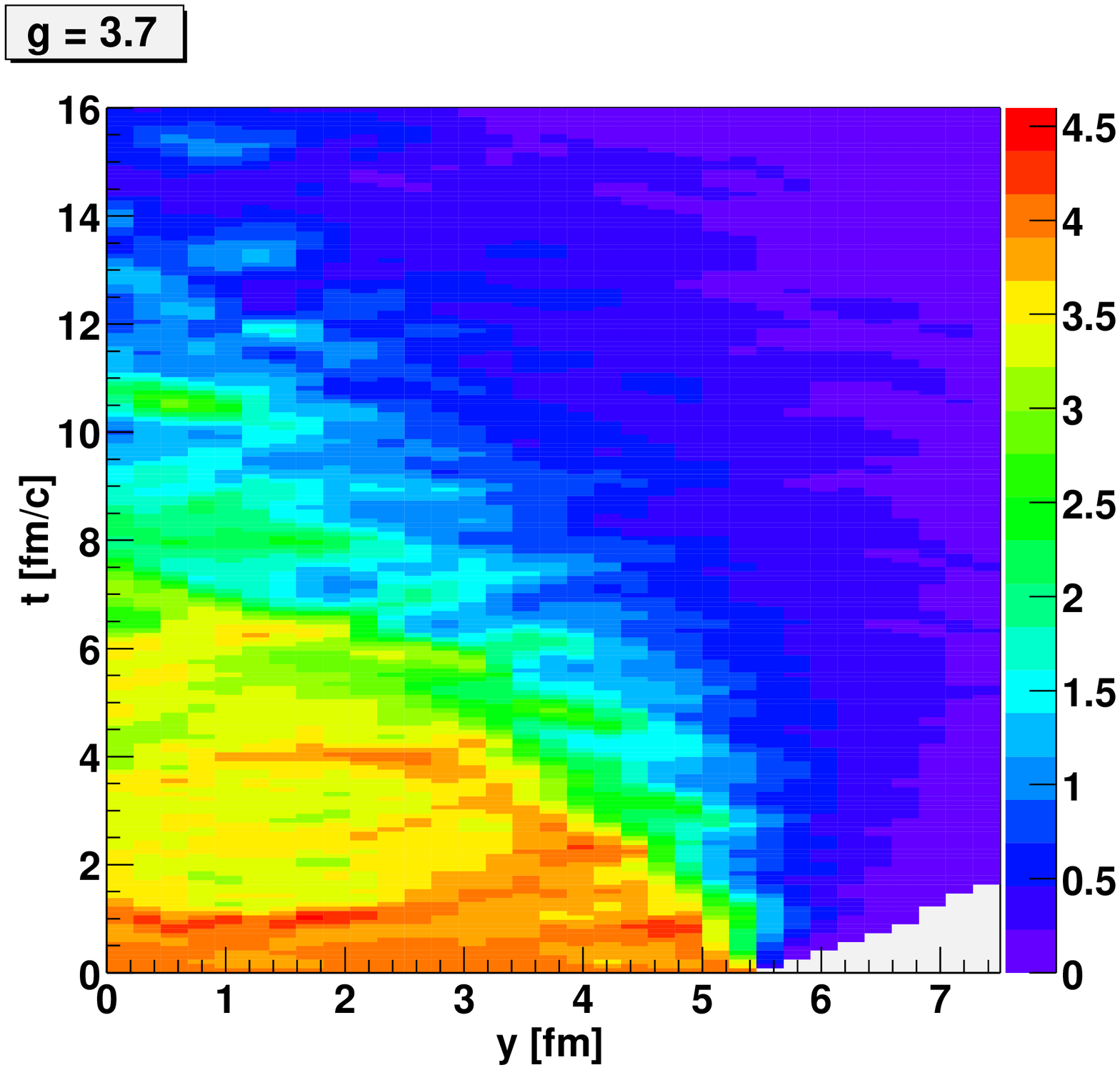,width=8cm}
\epsfig{figure=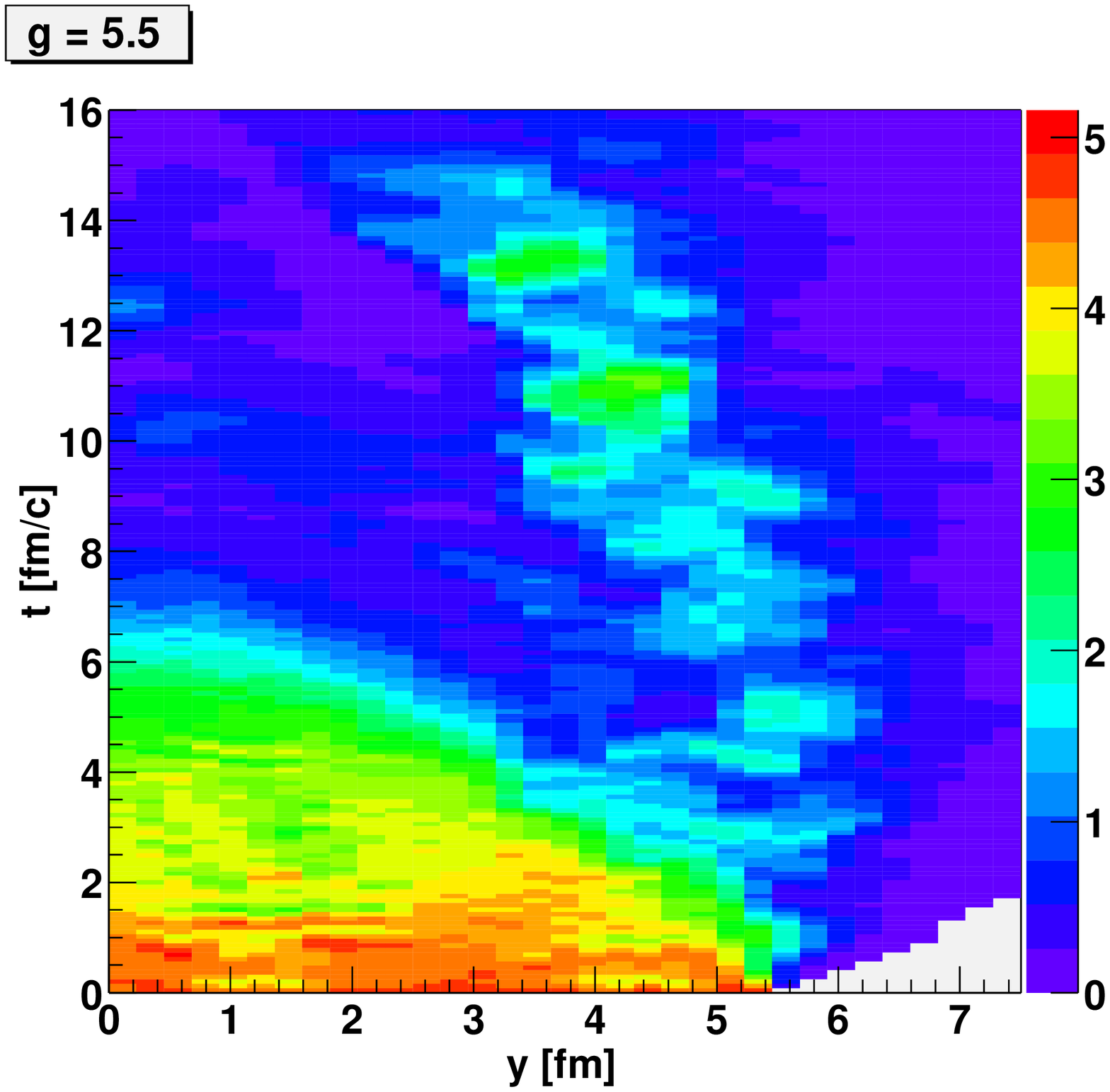,width=8cm}
\caption{Space-time evolution of the energy density along the $y$-axis
at $x=z=0$.} 
\label{compare_ey}
\end{figure}

The time evolution of the local rest-frame energy density $e$ of the quarks 
is shown in Figs.~\ref{compare_ex},~\ref{compare_ey}.
Again we see large-scale structures
for the first order phase transition ($g=5.5$), while
the energy density is rather homogeneous on large time and
distance scales if the expansion trajectory goes through the
critical endpoint. For the first-order transition, quarks
can be ``trapped'' in droplets with $\sigma\sim0$ (the
minimum of the effective potential where the symmetry is restored)
because the mass barrier can keep them from escaping.
The droplet of high-density matter at $x\simeq3$~fm and
$t\simeq12-16$~fm/c can easily be associated to the region of
nearly vanishing chiral scalar field from Fig.~\ref{compare_phix}.
Eventually, that region must perform the transition to the
symmetry broken state, either by a strong thermal fluctuation
or when reaching the spinodal point. At the spinodal, the system
is as far from local thermal equilibrium as it can get, and the
``roll-down'' of the order parameter field to the global minimum
of the potential can influence the collective expansion of
the quark fluid.  

Note that the energy density at the center drops
more rapidly for the first-order transition than near the critical endpoint.
This has consequences for the build-up of azimuthally asymmetric flow,
as we shall discuss below.

\begin{figure}[h!]
\centering
\epsfig{figure=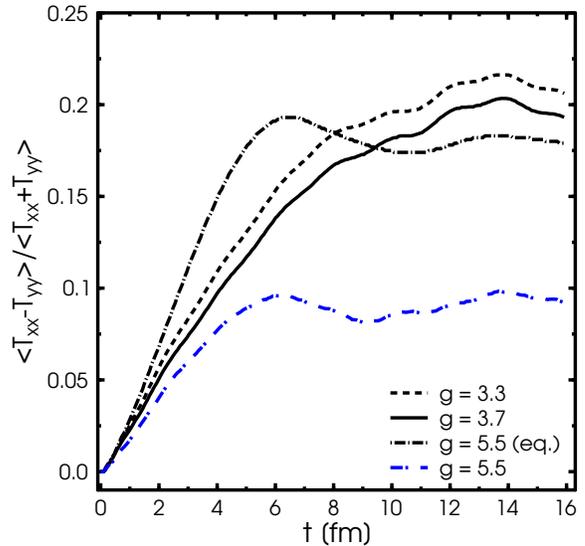,width=8cm}
\caption{Time evolution of the momentum anisotropy for a cross over
($g=3.3$), a second-order phase transition at the critical point
($g=3.7$), and a first order phase transition ($g=5.5$) in or out of
equilibrium.} 
\label{tii}
\end{figure}

Fig.~\ref{tii} depicts the time evolution of the 
azimuthal momentum anisotropy~\cite{KolbSollfrank}
\be
\epsilon_p=\frac{\langle T_{xx}-T_{yy}\rangle}{\langle T_{xx}+T_{yy}\rangle}~,
\ee
where the averages of the stress-energy tensor of the fluid
are taken at fixed time:
\be
\langle T_{ij}\rangle(t) \equiv \int d^3x \, T_{ij}(t,\vec{x})~.
\ee
Also, we average over a few initial field configurations, which
gave similar results for $\epsilon_p(t)$, though.

For the above-mentioned initial conditions
the energy-momentum tensor is symmetric, and so
$\epsilon_p=0$ at $t=0$ (this might be different in more realistic
treatments~\cite{ini_asym}). 
Pressure gradients in $x-$ and $y-$ directions are
different, though. Therefore, the acceleration of the fluid is
stronger in the reaction ($x-z$) plane than out of plane, leading
to a nonzero azimuthal asymmetry $\epsilon_p>0$ at times $t>0$.
The asymmetry first grows nearly linearly with time but saturates when
the asymmetry of the
energy density and of the pressure gradients becomes small. As
explained above (Figs.~\ref{compare_ex},~\ref{compare_ey}), this
happens earlier for a first-order transition than for trajectories
near the chiral critical endpoint. This is then reflected in the final
value of $\epsilon_p$. We stress that the more rapid saturation of
the azimuthal asymmetry in case of a first-order transition is not in
contradiction to the fact that hot (high-energy density) ``droplets''
survive for rather long times, as seen in the figures. Rather, such
``droplets'' typically turn out to be more or less rotationally
symmetric, or at least exhibit deformations which are uncorrelated
to the reaction plane (the $x-z$ plane in our case). Thus, they tend
to reduce the average azimuthal asymmetry of the energy-momentum
tensor.

For comparison, in Fig.~\ref{tii} we also show the result for an
equilibrium first-order phase transition. Here, the equations of
motion for the chiral fields, eqs.~(\ref{EulerLagrange}) are not
solved but rather the $\sigma$-field is required to populate the
(global) minimum of the effective potential,
\bea
\frac{\delta V_{\rm eff}}{\delta \sigma} &=& 0~,\\
\frac{\delta^2 V_{\rm eff}}{\delta \sigma^2} &>& 0~.
\eea
That is, the chiral field is in equilibrium with the quark-antiquark
fluid and does not exhibit any explicit space-time dependence.
At $T_c$, where two degenerate minima exist, one performs the usual
Maxwell-Gibbs construction to determine the fractions of the total
volume occupied by matter in the symmetric and the broken symmetry
phases, respectively. Evidently, for the above-mentioned initial
conditions the equilibrium phase transition
leads to nearly the same azimuthal asymmetry of $T^{ij}$ as for the
cross over. Therefore, it is indeed the non-equilibrium real-time dynamics
(field fluctuations over the free-energy barrier)
that is responsible for the observed reduction of $\epsilon_p$ in the
regime of first-order chiral phase transitions.

\section{Summary and Outlook} \label{outlook}
In summary, we have introduced a simple phenomenological description
of the non-equilibrium real-time dynamics of the chiral phase transition
in an expanding (relativistic) fluid of quarks. More precisely, we coupled
the linear sigma model,
which describes the dynamics of the long-wavelength modes of the chiral
order parameter field, to the hydrodynamical evolution of a system of quarks.
The chiral field(s) evolve according to the finite-temperature effective
potential that is generated by integrating out the quarks from the
Lagrangian; in turn, the field(s) determine the effective quark mass
(i.e.\ the equation of state of the quark fluid) dynamically.

The above model exhibits a first order phase transition for large
$g$, which is the quark-field coupling constant. The line of first order
transitions ends in a critical point when $g$ is lowered, i.e.\ the
transition turns into a cross over for smaller couplings. Thus, by varying $g$
one can qualitatively compare the hydrodynamic expansion pattern of the
quark fluid for dynamical trajectories that cross the line of first
order transitions to that obtained in the cross over regime.

We have obtained numerical solutions in 3+1 space-time dimensions, using
simple initial conditions that might be appropriate for relativistic heavy-ion
collisions. The hydrodynamical expansion pattern clearly depends on
the structure of the effective potential. For trajectories in the cross
over regime or near the critical endpoint the overall bulk dynamics is
found to be rather ``smooth'', in that the space-time distribution of
the energy density of the fluid is not affected very much by the
fluctuations of the order parameter field. In the absence of a latent
heat, the energy density can not jump much between regions where the
field amplitude is different.

In contrast, if the effective potential exhibits a barrier between the
symmetry restored and broken phases, respectively, we do see that
large-scale structures are formed dynamically, e.g.\ ``droplets'' of
the symmetric phase may survive for rather long times before becoming
mechanically unstable (at the spinodal). In that sense, the overall time
scale is longer for trajectories crossing the line of first order
transitions. Nevertheless, typically such structures are not correlated
to the reaction plane; thus, the direct correspondence of
spatial anisotropies in the initial condition to
momentum-space anisotropies in the final state predicted by
{\em equilibrium} hydrodynamics (that is,
when the phase transition is not treated
dynamically but modelled by a Maxwell-Gibbs construction) is weakened. 
For example, we find much smaller momentum-space
anisotropy for a dynamical
first-order transition than for a trajectory through the
chiral critical endpoint (for the {\em same initial condition}).
This could be a very useful prediction with regard to the experimental
search for the chiral critical endpoint of QCD in heavy-ion collisions
at the BNL-AGS, the CERN-SPS and the envisaged new GSI heavy-ion accelerator.
Until now, experiments focused on fluctuation observables, but
inclusive observables usually are much easier to analyze accurately.

\begin{figure}[h!]
\centering
\epsfig{figure=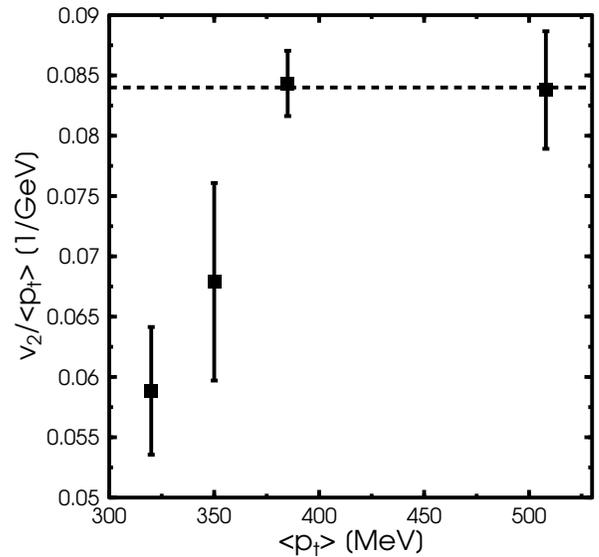,width=8cm}
\caption{Excitation function of $v_2/\langle p_{\rm t}\rangle$ 
of negatively charged particles
in mid-central collisions from top AGS to RHIC energy. Data for
$v_2$ are taken from Fig.~24 in~\protect\cite{na49_v2} 
and $\langle p_{\rm t}\rangle$
from~\protect\cite{pt_ags,pt_40,pt_158,pt_star}.}
\label{v2pt}
\end{figure}

In the future, we intend to scrutinize other inclusive
observables as to
their sensitivity to non-equilibrium effects from phase transitions.
Of course, there is also plenty of room to improve on the model in order
to obtain more quantitative predictions. The present paper represents a
first step towards an actual real-time description of the chiral phase
transition on either side of the critical endpoint in expanding
relativistic fluids with realistic 3+1d geometries.\\[1cm]
{\bf Note added:}
After this manuscript was submitted for publication the NA49 
collaboration published the elliptic flow at 
$E_{\rm lab}=40 A $~GeV~\cite{na49_v2}. 
From Fig.~24 of that publication, the dependence of $v_2$ on $\log\sqrt{s}$
is approximately linear. However, the ''natural'' scale for $v_2$ is
set by $\langle p_{\rm t}\rangle$, not $\log\sqrt{s}$, as pointed out
by Snellings~\cite{snellings}. Indeed, at high energies the differential
$v_2(p_{\rm t})$
of charged hadrons is approximately proportional to $p_{\rm t}$, such that 
the averaged $v_2\propto\langle p_{\rm t}\rangle$. 
In fact, for mid-central collisions $v_2$ increases from $\approx 3\%$ 
at top SPS energy ($\sqrt{s}=18 A$~GeV) to $\approx 4.5\%$ at RHIC energy
($\sqrt{s}=130 A$~GeV). When scaled by the average transverse momentum, though,
the elliptic flow in that energy regime is nearly constant~\cite{snellings}.
To scrutinze deviations from the ''natural'' scaling 
$v_2\propto\langle p_{\rm t}\rangle$, we plot the excitation function of 
$v_2/\langle p_{\rm t}\rangle$ in Fig.~\ref{v2pt}.

One observes that, as already mentioned above, the data is
compatible with no energy (or $\langle p_{\rm t}\rangle$) dependence
above top SPS energy. Clearly, there is
a systematic drop of $v_2$ {\em relative to}
$\langle p_{\rm t}\rangle$ towards lower energies. For instance, at
$E_{\rm lab}=40A$~GeV corresponding to $\langle p_{\rm
t}\rangle\approx350$~MeV, $v_2/\langle p_{\rm t}\rangle$ is lower by
about
two standard deviations than at higher energies. Qualitatively, this 
interesting behavior is similar to the reduction of the azimuthal 
momentum asymmetry, predicted above, 
caused by crossing the second order critical point
into the regime of first order phase transitions. Additional studies
of ``conventional'' non-equilibrium effects unrelated to a phase
transition are certainly required, however, before firm 
conclusions can be drawn.


\acknowledgements
We thank T.\ Kodama, I.\ Mishustin, O.~Scavenius and R.\ Snellings
for helpful discussions,
and D.~Rischke for a critical reading of the manuscript as
well as for communicating his 3+1d SHASTA code. K.P.\ and H.S.\ gratefully
acknowledge support by GSI, DFG, DESY, the Bergen 
Computational Physics Laboratory (Project No.\ 43) and the 
Frankfurt Center for Scientific Computing.
A.D.\ was partly supported by the U.S.\ Department of Energy
under contract No.\ DE-AC02-98CH10886 and by the German Ministry for
Education and Research (BMBF).

\begin{appendix}

\section{Subtracting Initial Fluctuations} \label{app_subtr}
In section~\ref{sec_effpot} we discussed our phenomenological {\em ansatz}
for the effective potential for the long wavelength modes of the chiral fields,
as generated by the heat bath of quarks. Formally, it is obtained from the
Lagrangian by integrating out the quarks to one loop.

Our main objective here is to study dynamically
fluctuations of the chiral order
parameter (or effects generated by those fluctuations) in the vicinity of the
chiral critical point as the system makes the transition to broken
chiral symmetry at low temperature. Thus, we have to allow for
primordial fluctuations of
the chiral fields, also. However, those fluctuations
of the fields at time $t=0$ will of course also contribute to the effective
potential and ``distort'' its shape. In order to restore our original choice
for $V_{\rm eff}$ from eq.~(\ref{T>0_potential}), and thus ensure the correct
dynamics for the long-wavelength modes, we have to introduce
appropriate subtractions.

The procedure is as follows. The scalar and pseudoscalar densities are given
by eqs.~(\ref{scal_dens}). They depend explicitly on the value of the fields
$\phi_a$, which we formally separate into short and long wavelengths:
\be \label{split}
\phi_a(x) = \left< \phi_a\right> + \delta\phi_a(x)~.
\ee
Here, $\langle\cdot\rangle$ denotes a spatial average over a volume large
enough for the fluctuations to average out:
\be
\left< \delta\phi_a\right> =0~.
\ee
The linear dimension of that volume will be given by the wavelength of the
soft modes of interest.

We now substitute~(\ref{split}) into eqs.~(\ref{scal_dens}) and perform
an expansion up to second order in $\delta\phi_a(x)$. We then perform the
averaging over the fluctuations with the distribution~(\ref{gaussdis})
and obtain
\bea
\left< \rho_s\right> &=& \rho_s(\left< \phi\right>) \nonumber\\
& + &\frac{1}{2} g \sigma \, d_q\int
\frac{{\rm d}^3 k}{(2\pi)^3}\biggl\{\biggr.
\nonumber\\
&&\sum\limits_a \left< \delta\phi_a^2 \right> \left[
- \frac{g^2 f(k)}{E^2 T} \left( \frac{T}{E} + f(k)e^{E/T}\right)\right.
\nonumber\\
&&\;\;\;+ \;\frac{g^4 \phi_a^2}{E^3 T^2} f(k)
\left( 2f^2(k)e^{2E/T} - f(k)e^{E/T}\right.
\nonumber\\
&&\;\;\; + \left.\left.3\frac{T}{E} f(k)e^{E/T} +3\frac{T^2}{E^2}\right)\right]
\nonumber\\
&&-  \left. 2 \left< \delta\sigma^2 \right>
\frac{g^2 f(k)}{E^2 T} \left( \frac{T}{E} + f(k)e^{E/T}\right)\right\}
\eea
Here, $\left< \delta\phi^2 \right>\equiv \sum_a
\left< \delta\phi_a\delta\phi_a
 \right>$ is the variance of the fluctuations in the initial condition,
i.e.\ at the initial time $t=0$, summed over internal quantum numbers.
We made use of the fact that the fluctuations~(\ref{gaussdis}) are diagonal
in internal space, i.e.\ $\left<\delta\phi_a\delta\phi_b\right>=0$ if
$a\neq b$.
The second term is the additional contribution seen by the long wavelength
modes $\left<\phi_a\right>$,
which is due to the fluctuations. To restore the original effective
potential, we have to subtract that term, i.e.\ redefine the scalar density
as
\bea\label{rhos_tayl}
\rho_s(\phi,T) &=& 
g \sigma \, d_q\int \frac{{\rm d}^3 k}{(2\pi)^3}
 \frac{1}{E} f(k) \nonumber\\
& - &\frac{1}{2} g \sigma \, d_q\int
\frac{{\rm d}^3 k}{(2\pi)^3}\biggl\{\biggr.
\nonumber\\
&&\sum\limits_a \left< \delta\phi_a^2 \right> \left[
- \frac{g^2 f(k)}{E^2 T} \left( \frac{T}{E} + f(k)e^{E/T}\right)\right.
\nonumber\\
&&\;\;\;+ \;\frac{g^4 \phi_a^2}{E^3 T^2} f(k)
\left( 2f^2(k)e^{2E/T} - f(k)e^{E/T}\right.
\nonumber\\
&&\;\;\; + \left.\left.3\frac{T}{E} f(k)e^{E/T} +3\frac{T^2}{E^2}\right)\right]
\nonumber\\
&&-  \left. 2 \left< \delta\sigma^2 \right>
\frac{g^2 f(k)}{E^2 T} \left( \frac{T}{E} + f(k)e^{E/T}\right)\right\}
\eea
This expression has to be substituted for $\rho_s$ on the right-hand-side
of the equation of motion~(\ref{EulerLagrange}). The subtracted term
cancels the ``distortion'' of the scalar density caused by using the local
values $\sigma(x)$, $\vec\pi(x)$ for the fields, rather than their
long-wavelength components $\langle\sigma(x)\rangle$, $\langle\vec\pi(x)
\rangle$.

Along the same lines one derives the following expressions for the
fluctuation-subtracted pseudo-scalar density, and for the pressure of
the quarks:
\bea\label{rhops_tayl}
\vec{\rho}_{\rm ps} (\phi,T)& =& 
 g \vec{\pi}\,  d_q\int \frac{{\rm d}^3 k}{(2\pi)^3}
 \frac{1}{E} f(k)\nonumber\\
& - &\frac{1}{2} g \vec{\pi} \, d_q\int
\frac{{\rm d}^3 k}{(2\pi)^3}\biggl\{\biggr.
\nonumber\\
&&\sum\limits_a \left< \delta\phi_a^2 \right> \left[
- \frac{g^2 f(k)}{E^2 T} \left( \frac{T}{E} + f(k)e^{E/T}\right)\right.
\nonumber\\
&&\;\;\;+ \;\frac{g^4 \phi_a^2}{E^3 T^2} f(k)
\left( 2f^2(k)e^{2E/T} - f(k)e^{E/T}\right.
\nonumber\\
&&\;\;\; + \left.\left.3\frac{T}{E} f(k)e^{E/T} +3\frac{T^2}{E^2}\right)\right]
\nonumber\\
&&-  \left. 2 \left< \delta\vec{\pi}^2 \right>
\frac{g^2 f(k)}{E^2 T} \left( \frac{T}{E} + f(k)e^{E/T}\right)\right\}
\nonumber\\
p(\phi,T) &=& d_q \int \frac{{\rm d}^3 k}{(2\pi)^3}
\, T \log\left(1+e^{-E/T}\right) \nonumber\\
&- & \frac{1}{2}  \sum\limits_a\left<\delta\phi_a^2\right> \left\{
- \frac{g^2 f(k)}{E}
\right.
\nonumber\\
&+&\left. \frac{g^4 \phi_a^2}{E^3 T} f(k)
\left[
T + E - E f(k)
\right]\right\}~.
\eea
One can verify that the identities $g\rho_s = - \delta p/\delta\sigma$ and
$g\rho_{ps} = - \delta p/\delta\vec\pi$ are satisfied, as it should be.

At fixed values for the fields, the energy density of the quarks at a
temperature $T$ is given by
\be
e(\phi,T) = T\frac{\partial p(\phi,T)}{\partial T} -p(\phi,T)~.
\ee
Using the expression for the fluctuation-subtracted pressure given above
one obtains
\bea
e(\phi,T) &= & d_q \int \frac{d^3k}{(2\pi)^3} E f(k) \nonumber\\
&-&   \frac{1}{2}  \sum\limits_a\left< \delta\phi_a^2 \right> \left\{ 
\frac{g^2 f(k)}{ET} \left[ T - E f(k) e^{E/T} \right] 
\right. \nonumber\\
&-& \frac{g^4 \phi_a^2}{E^3 T} f(k)
\left[
T + E - E f(k)
\right] 
\nonumber\\
&+& \left. \frac{g^4 \phi^2_a}{ET^2} f^2(k) e^{E/T} \tanh\left(\frac{E}{2T}\right) \right\}
\eea

The source term (\ref{source_field}) changes due to the fluctuations and
one has to use the modified scalar~(\ref{rhos_tayl}) and
pseudoscalar~(\ref{rhops_tayl}) densities, respectively.


To second order in the fluctuations, the self-interaction of the chiral fields
is renormalized as
\bea
U(\phi_a) &=& \frac{\lambda ^{2}}{4}\left(\sigma ^{2}+\vec{\pi} ^{2} -
{\it v}^{2}\right)^{2}-h_q\sigma -U_0 \nonumber\\
&-&   \frac{1}{2} \sum\limits_a\left< \delta\phi_a^2 \right> 
\lambda^2\left( 2\phi_a^2 + \sigma ^{2}+\vec{\pi} ^{2} - {\it v}^{2} \right)
\eea
The above expressions for $U(\phi_a)$,
$\rho_s$, $\vec{\rho}_{ps}$, $e$, and $p$ are
to be used in the equations of motion for the chiral
fields~(\ref{EulerLagrange}), in the stress-energy tensor of the quark
fluid~(\ref{idhyd_SET}), and in the source term for its divergence
$S^\nu$~(\ref{source_field},\ref{div_Tmunu_phi}).
We point out that we subtract those quantities for the contribution from
initial fluctuations of $\phi$ only up to second order in $\delta\phi$. We can
therefore not employ initial conditions with very large
local fluctuations about the mean field.
\end{appendix}

\end{document}